\documentclass[prd,aps,nofootinbib,showpacs,preprintnumbers,showkeys]
{revtex4}
\usepackage{graphicx,epsf,amsmath,amsfonts,amssymb,amsbsy}
\newcommand{\ds}{\displaystyle}
\newcommand{\vev}[1]{\langle#1\rangle}
\newcommand{\mat}{\left ( \begin{array}}
\newcommand{\emat}{\end{array} \right )}
\newcommand{\vect}{\left ( \begin{array}{c}}
\newcommand{\evect}{\end{array} \right )}

\preprint{HU-EP-06/44}

\begin{document}
\title{Diquarks in the color--flavor locked phase of dense quark
matter}
\author{D.~Ebert$^{\ast}$ and K.~G.~Klimenko$^{\dagger,\ddagger}$}
\affiliation{$^{\ast}$ Institut f\"ur Physik,
Humboldt-Universit\"at zu Berlin, 12489 Berlin, Germany}
\affiliation{$^{\dagger}$ Institute
for High Energy Physics, 142281, Protvino, Moscow Region, Russia}
\affiliation{$^{\ddagger}$ Dubna University (Protvino branch),
142281, Protvino, Moscow Region, Russia}

\begin{abstract}
Diquark excitations of dense quark matter 
are considered in the framework of the Nambu -- Jona-Lasinio model
with three types of massless quarks in the presense of a quark number
chemical potential $\mu$. We investigate the effective action
of meson- and diquark fields at sufficiently high values of $\mu$,
where the color--flavor locked (CFL) phase is realized, and prove the
existence of NG-bosons in the sector of pseudoscalar diquarks.
In the sector of scalar diquarks an additional NG-boson is found,
corresponding to the spontaneous breaking of the U(1)$_B$ baryon
symmetry in the CFL phase. Finally, the existence of massive scalar
and pseudoscalar diquark excitations is demonstrated.
\end{abstract}


\maketitle

\section{Introduction}

It is well-known that at sufficiently high baryon densities massless
three-flavor QCD is in the so-called color -- flavor locked (CFL)
phase \cite{alford1,alford2}. In this phase the original
SU(3)$_L\times$SU(3)$_R\times$SU(3)$_c\times$U(1)$_B$ symmetry of QCD
is spontaneously broken down to the diagonal SU(3)$_{L+R+c}$
subgroup. Correspondingly, seventeen massless excitations must appear
in the mass spectrum of the theory. Eight of them might be used to
ensure a mass of gluons by the Anderson -- Higgs mechanism. The
properties of the remaining nine, one scalar and eight pseudoscalar
Nambu -- Goldstone (NG) bosons, as well as other collective modes of
the CFL phase were studied already in the framework of 
weak-coupling QCD \cite{rho,miran,gusynin}.

It is clear that a weak-coupling QCD analysis of the color
superconductivity phenomena, including the CFL one, can only be
trusted at rather high baryon densities or, correspondingly, for
values of the quark number chemical potential $\mu>>1$ GeV
\cite{shuster}. At moderate values of $\mu\sim 500$ MeV, where 
weak-coupling QCD is not applicable, the description of color
superconductivity is more adequate in the framework of effective
theories for the low energy QCD region. In particular, since massless
excitations might play an important role in different transport
phenomena, a chiral effective meson theory for the pseudoscalar
NG-bosons of the CFL phase was also constructed
\cite{casalbuoni,andersen} (see also the recent review on this topic
\cite{schafer} and references therein).

Another effective quark model approach to low density QCD is based on
the Nambu -- Jona-Lasinio (NJL) type of models. Since the NJL model
contains the microscopic quark degrees of freedom, it is more
fundamental and preferable, especially for the investigation of
dynamical processes in dense baryonic matter, than chiral meson
theories. On the other hand, it is also more suitable for the
description of physics at low baryon densities than weak-coupling 
QCD. In particular, in the three-flavor NJL model the CFL effect was
already considered, e.g., in \cite{nebauer} (see also the review
\cite{buballa}), where some aspects of the phase structure of dense
quark matter were discussed, including the influence of the $s$-quark
bare mass, color- and electric charge neutrality conditions, external
magnetic field, etc. However, in spite of the fact that the lightest
bosons may play an essential role e.g. in the cooling processes of
neutron stars, up to now only few attention was paid to the
consideration of the CFL ground state bosonic excitations, i.e.
mesons and diquarks, and their dynamics in the framework of NJL
models (see, however, the papers \cite{reddy,iida}, where the
properties of the massless NG boson, corresponding to the spontaneous
breaking of the baryon U(1)$_B$ symmetry in the CFL phase, were
considered). In contrast, the properties of $\pi$-mesons and
diquarks, surrounded by color superconducting quark matter, were
already discussed in the framework of the two-flavor NJL model
\cite{bekvy,eky1,eky2,he,hashimoto}.
 
In the present paper we are going to study meson and diquark
excitations of the CFL ground state in the framework of the
massless three-flavor NJL model. As in \cite{bekvy,eky1,eky2}, our
consideration is based on the effective action, which is a generating
functional for one-particle irreducible Green functions. They permit
to get informations about the masses of bosonic excitations of the
CFL phase. In the first step we will rederive the well--known result
that in the normal phase of quark matter the NG bosons are just
pseudoscalar mesons. Next, it will be shown that in the CFL phase the
NG bosons are scalar- and pseudoscalar diquark excitations. Finally,
we will demonstrate that in the CFL phase there appear massive scalar
and pseudoscalar diquarks which are composed into a charged triplet
and singlet as well as a neutral singlet of the SU(3) group.

\section{NJL model and effective meson-diquark action}

Let us consider the following NJL model with three massless quark
flavors
\begin{eqnarray}
&&  L=\bar q\Big [\gamma^\nu i\partial_\nu+  \mu\gamma^0\Big
]q+ G_1\sum_{a=0}^8\Big [(\bar
  q\tau_aq)^2+(\bar qi\gamma^5\tau_a q)^2\Big ]+\nonumber\\
&+&G_2\!\!\!\sum_{A=2,5,7}\sum_{A'=2,5,7}\Big\{
[\bar q^Ci\gamma^5\tau_A\lambda_{A'}q]
[\bar qi\gamma^5\tau_A\lambda_{A'} q^C]
+[\bar q^C\tau_A\lambda_{A'}q]
[\bar q\tau_A\lambda_{A'} q^C]\Big\}.
  \label{1}
\end{eqnarray}
In (\ref{1}), $\mu\geq 0$ is the quark number chemical potential,
which is the same for all quark flavors, $q^C=C\bar q^t$, $\bar
q^C=q^t C$ are charge-conjugated spinors, and
$C=i\gamma^2\gamma^0$ is the charge conjugation matrix (the symbol
$t$ denotes the transposition operation). The quark field
$q\equiv q_{i\alpha}$ is a flavor and color triplet as
well as a four-component Dirac spinor, where $i,\alpha=1,2,3$.
(Roman and Greek indices refer to flavor and color
indices, respectively; spinor indices are omitted.) Furthermore,
we use the notations  $\tau_a,\lambda_a$ for Gell-Mann matrices in
the flavor and color space, respectively ($a=1, ...,8)$; $\tau_0
=\sqrt{\frac{2}{3}}\,1\hspace{-1.2mm}1$ is proportional to the unit
matrix in the flavor space. Clearly, the Lagrangian (\ref{1}) as a
whole is invariant under transformations from the color group
SU(3)$_c$. In addition, it is symmetric under the chiral group
SU(3)$_L\times$SU(3)$_R$ (chiral transformations act on the flavor
indices of quark fields only) as well as under the baryon-number
conservation group U(1)$_B$ and the axial group U(1)$_A$. 
\footnote{In a more realistic case, the additional
`t Hooft six-quark interaction term should be taken into account in
order to break the axial U(1)$_A$ symmetry \cite{nebauer}. However,
in the present consideration we omit the `t Hooft term, for
simplicity.}

The linearized version of the Lagrangian (\ref{1}) contains
collective bosonic fields and looks like
\begin{eqnarray}
\tilde L\ds &=&\bar q\Big [\gamma^\nu i\partial_\nu +\mu\gamma^0
-\sigma_a\tau_a-i\gamma^5\pi_a\tau_a\Big ]q
-\frac{1}{4G_1}\Big [\sigma_a\sigma_a+ \pi_a\pi_a\Big ]-
 \frac1{4G_2}\Big [\Delta^{s*}_{AA'}\Delta^s_{AA'}+
 \Delta^{p*}_{AA'}\Delta^p_{AA'}\Big ]
 \nonumber\\
&-&\frac{\Delta^{s*}_{AA'}}{2}[\bar
q^Ci\gamma^5\tau_A\lambda_{A'} q]
-\frac{\Delta^s_{AA'}}{2}[\bar q i\gamma^5\tau_A\lambda_{A'}
q^C]-\frac{\Delta^{p*}_{AA'}}{2}[\bar 
q^C\tau_A\lambda_{A'} q]
-\frac{\Delta^p_{AA'}}{2}[\bar q\tau_A\lambda_{A'} q^C],
\label{2}
\end{eqnarray}
where here as well as in the following the summation over  repeated
indices $a=0,...,8$ and $A,A'=2,5,7$ is implied.  Lagrangians
(\ref{1}) and (\ref{2}) are equivalent which simply follows from
the equations of motion for the bosonic fields
\begin{eqnarray}
\sigma_a(x)=-2G_1(\bar
q\tau_aq),~~~\Delta^s_{AA'}(x)\!\!&=&\!\!-2G_2(\bar
q^Ci\gamma^5\tau_A\lambda_{A'}q),~~~
\Delta^{s*}_{AA'}(x)=-2G_2(\bar qi\gamma^5\tau_A\lambda_{A'} q^C),
\nonumber\\\pi_a(x)=-2G_1(\bar
qi\gamma^5\tau_a q),~~~ \Delta^p_{AA'}(x)\!\!&=&\!\!-2G_2(\bar
q^C\tau_A\lambda_{A'}q),~~~
\Delta^{p*}_{AA'}(x)=-2G_2(\bar q \tau_A\lambda_{A'} q^C).
\label{3}
\end{eqnarray}
One can easily see from (\ref{3}) that the mesonic fields $\sigma_a
(x),\pi_a (x)$ are real quantities, i.~e. $(\sigma_a(x))^\dagger=
\sigma_a(x),~~(\pi_a(x))^\dagger=\pi_a(x)$ (the superscript symbol
$\dagger$ denotes the hermitian conjugation), whereas all diquark
fields $\Delta^{s,p}_{AA'}(x)$ are complex ones, i.~e.
\[
(\Delta^s_{AA'}(x))^\dagger=\Delta^{s*}_{AA'}(x),~~~~~
(\Delta^p_{AA'}(x))^\dagger=\Delta^{p*}_{AA'}(x).
\]
Moreover, $\Delta^s_{AA'}(x)$ and $\Delta^p_{AA'}(x)$ are scalars
and
pseudoscalars, correspondingly. 

Let us introduce the flavor group SU(3)$_f$=SU(3)$_{L+R}$, which is
the diagonal subgroup of the chiral group. Then, all scalar diquarks
$\Delta^s_{AA'}(x)$ form an $(\bar 3_c,\bar 3_f)$-multiplet of the
SU(3)$_c\times$SU(3)$_f$ group, i. e. they are a color and flavor
antitriplet. The same is true for pseudoscalar diquarks
$\Delta^p_{AA'}(x)$ which are also the components of an $(\bar
3_c,\bar 3_f)$-multiplet of the SU(3)$_c\times$SU(3)$_f$ group.
Evidently, all diquarks $\Delta^{s,p}_{AA'}(x)$ have the same
nonzero baryon charge. All the real $\sigma_a(x)$ and $\pi_a(x)$
fields are color singlets. Moreover, the set of scalar $\sigma
_a(x)$-mesons is decomposed into a direct sum of the singlet and
octet representations of the diagonal flavor group SU(3)$_f$. The
same decomposition into multiplets is true for the set of all
pseudoscalar $\pi_a(x)$-mesons. Clearly, in this case the octet is
constructed from three pions ($\pi^{\pm}$ and $\pi^0$), four kaons
($K^0$, $\bar K^0$ and $K^\pm$) and the eta-meson ($\eta_8$),
whereas the singlet ($\eta_0$) corresponds to the
$\eta^{\,\prime}$-meson. 

Next, in order to use the Nambu--Gorkov formalism, we put the quark
fields and their charge conjugates together into a bispinor
$\Psi=\binom q{q^C}$ so that the Lagrangian (\ref{2}) takes the
compact form
\begin{equation}
\label{4}
\tilde L\ds =
-\frac{1}{4G_1}\Big [\sigma_a\sigma_a+ \pi_a\pi_a\Big ]-
 \frac1{4G_2}\Big [\Delta^{s*}_{AA'}\Delta^s_{AA'}+
 \Delta^{p*}_{AA'}\Delta^p_{AA'}\Big ]+
\frac 12\bar\Psi Z\Psi,
\end{equation}
where $Z$ is the $2\times 2$-matrix in the Nambu--Gorkov space,
\begin{equation}
Z=\left (\begin{array}{cc}
 D^+, & - K\\
- K^*, & D^-
\end{array}\right ),
\label{5}
\end{equation}
and the following notations are adopted
\begin{eqnarray}
&&D^+=i\gamma^\nu\partial_\nu+\mu\gamma^0-\Sigma,~~~~~~~
\Sigma=\tau_a\sigma_a+ i\gamma^5\pi_a\tau_a,~~~~~~~
K=(\Delta^p_{AA'}+i\Delta^s_{AA'}\gamma^5)\tau_A\lambda_{A'},
\nonumber\\&&
D^-=i\gamma^\nu\partial_\nu-\mu\gamma^0-\Sigma^t,~~~~~~
\Sigma^t=\tau_a^t\sigma_a+ i\gamma^5\pi_a\tau^t_a,~~~~~~
K^*=(\Delta^{p*}_{AA'}+i\Delta^{s*}_{AA'}\gamma^5)\tau_A
\lambda_{A'}.
\label{6}
\end{eqnarray}
Now, integrating out the quark fields in the partition function with
the Lagrangian (\ref{4}) (this procedure is presented in detail in
\cite{eky2}), we obtain the effective action for the considered NJL
model in the one-fermion loop approximation,
\begin{equation}
{\cal S}_{\rm {eff}}(\sigma
_a,\pi_a,\Delta^{s,p}_{AA'},\Delta^{s,p*}_{AA'})
=-\int d^4x\left[\frac{\sigma^2_a+\pi^2_a}{4G_1}+
\frac{\Delta^s_{AA'}\Delta^{s*}_{AA'}+\Delta^p_{AA'}\Delta^{p*}_{AA'}
}{4G_2}\right]-\frac i2{\rm Tr}\ln Z. 
\label{7}
\end{equation}
Besides of an evident trace over the two-dimensional Nambu--%
Gorkov matrix, the Tr-operation in (\ref{7}) stands for calculating
the trace in spinor, flavor, color as well as four-dimensional
coordinate spaces, correspondingly. Let us suppose that parity is
conserved so that all pseudoscalar diquark and meson fields have
zero ground state expectation values, i.~e.
$\vev{\Delta^p_{AA'}(x)}=0$ and $\vev{\pi_a(x)}=0$. Furthermore,
since at zero $s$-quark mass, $m_s=0$, only the competition between
the normal quark matter phase and the CFL one is relevant to the
physics of dense QCD (see, e.g., Fig. 1 in \cite{alford2}, where the
corresponding phase portrait for QCD with two massless $u,d$-quarks
at zero temperature is presented in terms of $\mu,m_s$), we permit in
the present consideration nonzero ground state expectation values
only for $\sigma_0(x)$ and $\Delta^s_{AA}(x)$ fields ($A=2,5,7$).
Namely, let $\vev{\sigma_0 (x)} \equiv\sigma_0^o,
~\vev{\Delta^{s}_{AA}(x)}\equiv\Delta$, $\vev{\Delta^{s*}_{AA}(x)}
\equiv\Delta^{*}$, where $A=2,5,7$, but other boson fields from
(\ref{3}) have zero ground state expectation values. In this case, if
$\Delta=0$, then quark matter is in the normal phase, where at
$\sigma_0^o\ne 0$ the ground state is an
SU(3)$_c\times$SU(3)$_f\times$U(1)$_B$-invariant one. If $\Delta\ne
0$, then the CFL phase is realized in the model, and the initial
symmetry is spontaneously broken down to SU(3)$_{L+R+c}$. Now, let
us make the following shifts of bosonic fields in (\ref{7}):
$\sigma_0 (x)\to\sigma_0(x)+\sigma_0^o$,
$\Delta^{s*}_{AA}(x)\to\Delta^{s*}_{AA} (x)+\Delta^*$,
$\Delta^{s}_{AA}(x) \to\Delta^{s}_{AA}(x)+\Delta$, where $A=2,5,7$,
and other bosonic fields remain unshifted. (Obviously, the new
shifted bosonic fields $\sigma_0 (x),\Delta^s_{AA}(x)$ etc, now
denote the (small) quantum fluctuations around the
mean values $\sigma_0^o,\Delta$ etc of mesons and diquarks rather
than the original fields (\ref{3}).) In this case 
\begin{equation}
Z\rightarrow\left (\begin{array}{cc}
D^+_o~, & -K_o\\
 -K_o^*~, & D^-_o
\end{array}\right )-\left (\begin{array}{cc}
\Sigma~, & K\\
 K^*~, & \Sigma^t
\end{array}\right )\equiv S_0^{-1}-\left (\begin{array}{cc}
\Sigma~, & K\\
 K^*~, & \Sigma^t
\end{array}\right ), 
\label{9}
\end{equation}
where $K_o,K^*_o, D^\pm_o,\Sigma_o,\Sigma^t_o$ are the corresponding
quantities (\ref{6}), in which all bosonic fields are replaced by
their own ground state expectation values, i.e. $\sigma_0(x)\to
\sigma_0^o$, $\pi_a(x)\to 0$,  $\Delta^{s}_{AA}(x)\to\Delta$,
$\Delta^{p}_{AA'}(x)\to 0$ etc, and $S_0$ is the quark propagator
matrix in the Nambu--Gorkov representation (its matrix elements
$S_{ij}$ are given in the Appendix). Then, expanding the obtained
expression into a Taylor-series up to second order of small bosonic
fluctuations, we have
\begin{equation}
{\cal S}_{\rm
{eff}}(\sigma_a,\pi_a,\Delta^{s,p}_{AA'},\Delta^{s,p*}_{AA'})=
{\cal S}_{\rm {eff}}^{(0)} +{\cal S}_{\rm  {eff}}^{(2)}
(\sigma_a,\pi_a,\Delta^{s,p}_{AA'},\Delta^{s,p*}_{AA'})+\cdots,
\label{10}
\end{equation}
where 
\begin{eqnarray}
 &&{\cal S}_{\rm {eff}}^{(0)}=-\int
 d^4x\left[\frac{\sigma^o_0\sigma^o_0}{4G_1}+
\frac{3|\Delta|^2}{4G_2}\right]+\frac i2{\rm
Tr}\ln \left (S_0\right )\equiv -\Omega(\sigma^o_0,\Delta,
 \Delta^{*})\int d^4x,   \label{11}\\
&&{\cal S}^{(2)}_{\rm {eff}}
(\sigma_a,\pi_a,\Delta^{s,p}_{AA'},\Delta^{s,p*}_{AA'})
 = -\int d^4x\left[\frac{\sigma^2_a+\pi^2_a}{4G_1}+
\frac{\Delta^s_{AA'}\Delta^{s*}_{AA'}+\Delta^p_{AA'}
\Delta^{p*}_{AA'}}{4G_2}\right]+\nonumber\\
&&~~~~~~~~~~~~~~~~~~~~~~+\frac i4{\rm
Tr}\left\{S_0\left (\begin{array}{cc}
\Sigma~, & K\\
 K^*~, & \Sigma^t
\end{array}\right )S_0\left (\begin{array}{cc}
\Sigma~, & K\\
 K^*~, & \Sigma^t
\end{array}\right )\right\}.
  \label{12}
\end{eqnarray}
Notice that the term linear in meson and diquark fields vanishes in
(\ref{10}) due to the gap equations (see below). In (\ref{11}) the
quantity $\Omega$ is the thermodynamic potential of the system. In
terms of $M=\sqrt{\frac 23}\sigma_0^o$ it looks like
\begin{eqnarray}
&&\Omega(M,\Delta,\Delta^*)=
\frac{3M^2}{8G_1}+\frac{3|\Delta|^2}{4G_2}-
8\int\frac{d^3q}{(2\pi)^3}\Big\{E_{\Delta}^++
E_{\Delta}^-\Big\}-
\int\frac{d^3q}{(2\pi)^3}\Big\{E_{2\Delta}^++
E_{2\Delta}^-\Big\},
\label{13}
\end{eqnarray}
where $(E_{\Delta}^\pm)^2=(E^\pm)^2+|\Delta|^2$, 
$(E_{2\Delta}^\pm)^2=(E^\pm)^2+4|\Delta|^2$, $E^\pm=E\pm\mu$,
$E=\sqrt{\strut\vec q^2+M^2}$. Starting from (\ref{13}), one can
find the gap equations $\partial\Omega /\partial\Delta^* =0$ and
$\partial\Omega /\partial M =0$, which supply us with the values
of $M,\Delta$ in the ground state of the system:
\begin{eqnarray}
&&\frac{\partial\Omega}{\partial M}
\equiv\frac{3M}{4G_1}-2M\!\int\!\frac{d^3q}{(2\pi)^3E}
\left\{\frac{4E^+}{E_{\Delta}^+}+
\frac{4E^-}{E_{\Delta}^-}+
\frac{E^+}{2E_{2\Delta}^+}+
\frac{E^-}{2E_{2\Delta}^-}\right\}=0,
\label{14}\\
&&\frac{\partial\Omega}{\partial\Delta^*}\equiv
\frac{3\Delta}{4G_2}-\Delta\!\int\!\frac{d^3q}{(2\pi)^3}
\left\{\frac{4}{E_{\Delta}^+}+
\frac{4}{E_{\Delta}^-} +\frac{2}{E_{2\Delta}^+}+
\frac{2}{E_{2\Delta}^-}\right\}=0.
\label{15}
\end{eqnarray}
Since the integrals in the right hand sides of (\ref{13})-(\ref{15})
are ultraviolet divergent, we regularize them as well as the other
three-dimensional divergent integrals below by implementing a cutoff
in the integration regions, $|\vec q|<\Lambda$. In all numerical
calculations below, we use the following values of the model
parameters (see, e.g., ref. \cite{buballa})
\begin{eqnarray}
&&\Lambda=602.3~{\rm MeV},~~~~ G_1\Lambda^2=2.319,~~~~G_2=3G_1/4.
\label{16}
\end{eqnarray}
Suppose that $\Delta$ is a real quantity. In this case, since the
thermodynamic potential $\Omega$ (\ref{13}) is an even function 
in both the $M$ and $\Delta$ variables, it is enough to study it only
in the region $\{M\ge 0,\Delta\ge 0\}$. Just for this region a graph
of $\Omega(M,\Delta)$ is presented in Fig. 1 at $\mu=\mu_c\approx
329$ MeV. It is clear from Fig. 1 that in this case the thermodynamic
potential has two global minimum points (GMP), $A$ and $B$. The
point $A$ with coordinates $(M_c,0)$, where $M_c\approx 355$ MeV,
lies on the $M$-axis and corresponds to the normal quark matter
phase, whereas the point $B$ with coordinates $(0,\Delta_c)$, 
$\Delta_c\approx 86$ MeV, is arranged on the $\Delta$-axis and
corresponds to the CFL phase of dense baryonic matter. (Evidently,
both $A$ and $B$ are solutions of the gap equations
(\ref{14})-(\ref{15}). In addition, as is seen from Fig. 1,
there are another solutions of the gap equations which, however,
correspond to a maximum- or saddle points of the thermodynamic
potential.) If $\mu<\mu_c$, the function $\Omega$ has only one
GMP of the $A$-type, i.e. it lies on the $M$-axis. So, in this case
the normal quark matter phase is realized in the model. In contrast,
if $\mu>\mu_c$, then the single GMP of the $\Omega$-function is
located on the $\Delta$-axis, i.e. it is of the $B$-type, and the CFL
phase occurs. The dependence of the coordinates (gaps) of the GMP on
the chemical potential is presented in Fig. 2. Since the GMP jumps at
$\mu=\mu_c$ from $A$ to $B$ (or vice versa), one may conclude that 
at $\mu=\mu_c$ a first order phase transition takes place in the
system.
\begin{figure}
 \includegraphics[width=0.45\textwidth]{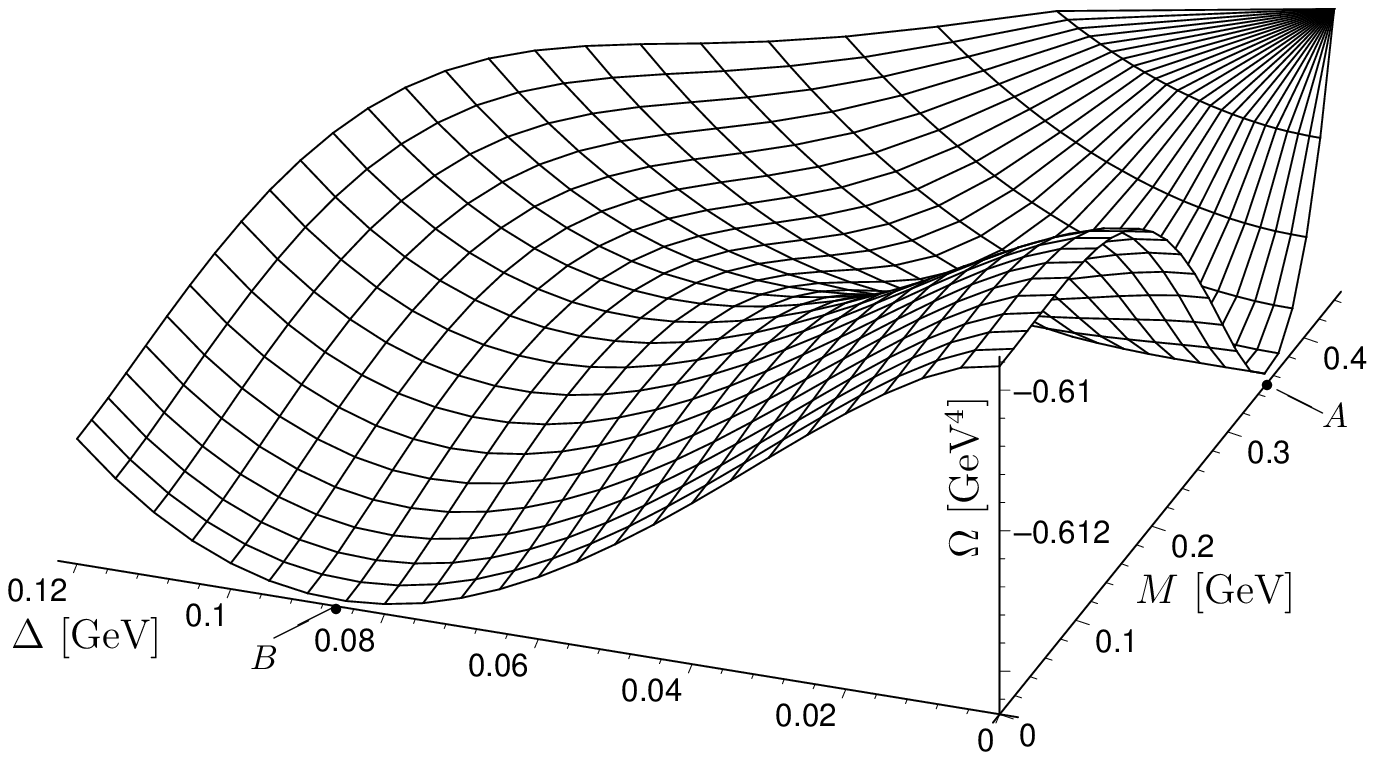}
 \hfill
 \includegraphics[width=0.45\textwidth]{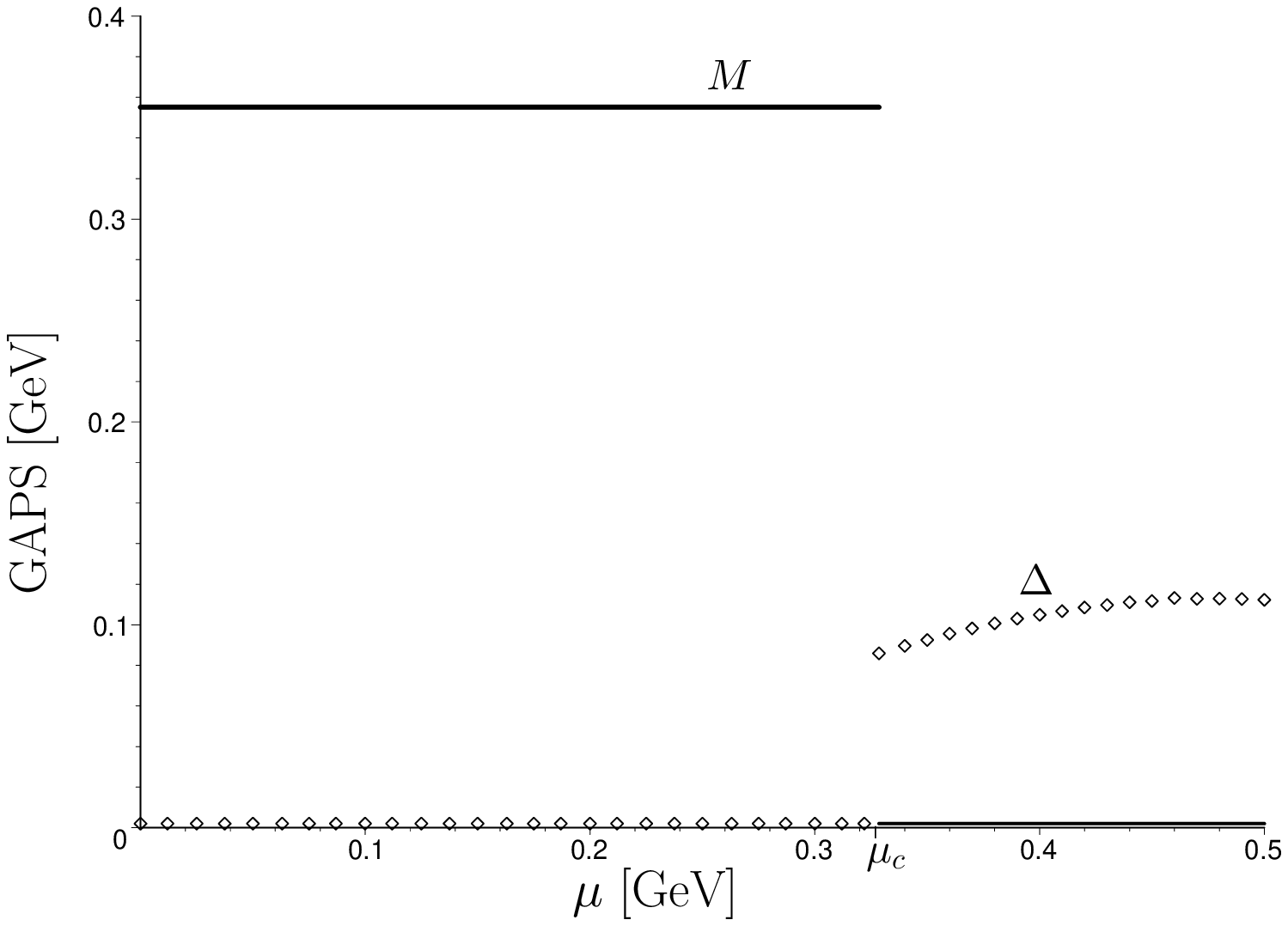}\\
\parbox[t]{0.45\textwidth}{
 \caption{The behaviour of the thermodynamic potential $\Omega$ vs
 $M$ and $\Delta$ at the critical value of the chemical potential 
 $\mu_c\approx 329$ MeV.} \label{fig:1} }
\hfill
\parbox[t]{0.45\textwidth}{
\caption{The coordinates $M$ and $\Delta$ (gaps) of the global
minimum point of the thermodynamic potential vs the chemical
potential $\mu$. Here $\mu_c\approx 329$ MeV.} \label{fig:2} }
\end{figure}

The effective action ${\cal S}_{\rm {eff}}^{(2)}$ in (\ref{12}) is
really a generating functional of the one-particle irreducible (1PI)
two-point Green functions of mesons and diquarks both in the normal
and CFL phases, namely 
\begin{eqnarray}
&& \Gamma_{XY}(x-y)=-\frac{\delta^2{\cal S}^{(2)}_{\rm eff}}{\delta
Y(y)\delta X(x)},
 \label{17}
\end{eqnarray}
where $X(x),Y(x)=\sigma_a(x),\pi_b(x),\Delta^{s,p}_{AA'}(x),
\Delta^{s,p*}_{BB'}(x)$. 
These Green functions are very useful, in particular, in determining
the dispersion relations and masses of particles. In the following,
we shall say that in the theory there is a {\it mixing} between two
different particles with corresponding fields $X(x)$ and $Y(x)$, if
their 1PI Green function $\Gamma_{XY}(x-y)$ is not equal to zero. By
analyzing the structure of the effective action (\ref{12}), it is
possible to show that in the NJL model (\ref{1}) with three massless
quarks, mesons and diquarks are not mixed both in the normal 
($\mu<\mu_c$) and CFL phases ($\mu>\mu_c$).
\footnote{Note, if some of the current quark masses are nonzero, then
there arises a mixing between mesons and diquarks in the CFL phase.
This effect is the analogy of the mixing between the $\sigma$-meson
and the scalar diquark in the color superconducting phase of a
two-flavor NJL model with nonzero masses of $u$- and $d$-quarks
\cite{eky1,eky2}.}
Moreover, each pseudoscalar $\pi_a(x)$ field (as well as scalar
$\sigma_b(x)$ one) does not mix with other meson fields, apart from
itself. So, in the normal phase (where $\Delta=0$ and $M\approx 355$
MeV) one can obtain the following expression $\overline{\Gamma}_
{\pi_i\pi_i}(p)$ for the Fourier transformation of the two-point 1PI
Green function of pseudoscalar mesons $\pi_i(x)$, taken in the rest
frame $p=(p_0,0,0,0)$:
\begin{eqnarray}
&&\overline{\Gamma}_{\pi_i\pi_i}(p_0)=p_0^2\int\frac{d^3q}{(2\pi)^3}
\cdot\frac{12}{E[p_0^2-4E^2]}.
 \label{18}
\end{eqnarray}
(In obtaining (\ref{18}), the gap equation (\ref{14}) was used in
order to eliminate the coupling constant $G_1$ from the final
expression.) Evidently, this expression turns into zero at
$p_0^2=0$. Since the relation (\ref{18}) is true for $i=0,1,...,8$,
it means that nine massless excitations, Nambu -- Goldstone (NG)
bosons, do exist in the pseudoscalar meson sector of the model in the
normal phase. This fact corresponds to a spontaneous symmetry
breaking in the normal phase down to the group
SU(3)$_c\times$SU(3)$_{f}\times$U(1)$_B$. 

Recall, that our initial NJL model (\ref{1}) is invariant under
the symmetry group
SU(3)$_L\times$SU(3)$_R\times$SU(3)$_c\times$ U(1)$_B\times$U(1)$_A$,
which contains an additional (nonphysical) axial subgroup U(1)$_A$.
So in the CFL phase, where the ground state is symmetric under the 
group SU(3)$_{L+R+c}$, an additional NG boson must appear in the mass
spectrum of the model giving altogether eighteen NG bosons.

Note that in the sector of the meson excitations of the CFL phase we
do not find any massless particle.

\section{Massless diquark excitations in the CFL phase}

A further consideration of the effective action (\ref{12}) shows that
in the CFL phase, where $M=0$ and $\Delta\ne 0$, scalar and
pseudoscalar diquarks are separated from each other. Moreover, the
eighteen pseudoscalar diquarks (nine $\Delta^p_{AA'}(x)$ and nine
$\Delta^{p*}_{AA'}(x)$ fields) may be divided into four sectors: 
$p(57,75)$, $p(25,52)$, $p(27,72)$ and $p(257)$. Each of the sectors
$p(AA',A'A)$, where $A\ne A'$, is composed of $\Delta^p_{AA'}(x)$,
$\Delta^{p*}_{AA'}(x)$, $\Delta^p_{A'A}(x)$ and $\Delta^{p*}_{A'A}(x)
$-diquarks, whereas the sector $p(257)$ is composed of six fields,
$\Delta^{p*}_{22}(x)$, $\Delta^{p*}_{55}(x)$, $\Delta^{p*}_{77}(x)$,
$\Delta^{p}_{22}(x)$, $\Delta^{p}_{55}(x)$, and $\Delta^{p}_{77}(x)$.
There is a mixing between diquarks from the same sector, however
fields from different sectors are not mixed. (The analogous situation
takes place for the set of scalar diquarks.) 

Partially, these mixing properties of the pseudoscalar diquarks are
explained by the ground state SU(3)$_{L+R+c}$ symmetry of the
CFL phase. Indeed, with respect to this group all pseudoscalar
diquarks are decomposed into a direct sum of the $\bar 3$ and $\bar
6$ multiplets. The fields $\Delta^p_{AA}(x)$ with $A=2,5,7$ are in
an antitriplet $\bar 3$, whereas all diquarks of the form
$\Delta^p_{A'A}(x)$ ($A'\ne A$) are in an antisixtet $\bar 6$. It is
clear from symmetry considerations that diquarks from $\bar 3$ do
not mix with diquarks from $\bar 6$ (here we use the terminology,
introduced just after (\ref{17})). Since the mixing between a
particle and corresponding antiparticle is allowed as a rule, we may
conclude that pseudoscalar diquarks from the sector $p(257)$ are
separated from the other six $\Delta^p_{A'A}(x)$- and six
$\Delta^{p*}_{A'A}(x)$-diquarks ($A'\ne A$). A further separation
between components of the $\bar 6$-multiplet and corresponding
antiparticles occurs on a dynamical basis, i. e. it is due to the
structure of the effective action (\ref{12}) and the quark propagator
matrix (\ref{121})-(\ref{124}). As a result, one can show that these
diquarks are divided into the three above-mentioned sectors,
$p(57,75)$, $p(25,52)$, $p(27,72)$.

Below we suppose that the gap $\Delta$ is a real positive number in
the CFL phase.

\subsection{The case of $p(AA',A'A)$ sectors}

Let us first study the mass spectrum of the excitations, e.g., in
the sector $p(57,75)$. The two-point 1PI Green functions of
pseudoscalar diquarks from this sector can be obtained from
(\ref{12}) by the relation (\ref{17}). In the rest frame, where
$p=(p_0,0,0,0)$, the Fourier transforms of these 1PI Green functions
form the following matrix:
\begin{equation}
\overline{\Gamma}_{57,75}(p_0)=\mat {cccc}
\overline{\Gamma}_{\Delta^p_{57}\Delta^{p}_{57}}(p_0)&
\overline{\Gamma}_{\Delta^p_{57}\Delta^{p*}_{57}}(p_0)&
\overline{\Gamma}_{\Delta^p_{57}\Delta^{p}_{75}}(p_0)&
\overline{\Gamma}_{\Delta^p_{57}\Delta^{p*}_{75}}(p_0)\\
\overline{\Gamma}_{\Delta^{p*}_{57}\Delta^p_{57}}(p_0)&
\overline{\Gamma}_{\Delta^{p*}_{57}\Delta^{p*}_{57}}(p_0)&
\overline{\Gamma}_{\Delta^{p*}_{57}\Delta^p_{75}}(p_0)&
\overline{\Gamma}_{\Delta^{p*}_{57}\Delta^{p*}_{75}}(p_0)\\
\overline{\Gamma}_{\Delta^p_{75}\Delta^{p}_{57}}(p_0)&
\overline{\Gamma}_{\Delta^p_{75}\Delta^{p*}_{57}}(p_0)&
\overline{\Gamma}_{\Delta^p_{75}\Delta^{p}_{75}}(p_0)&
\overline{\Gamma}_{\Delta^p_{75}\Delta^{p*}_{75}}(p_0)\\
\overline{\Gamma}_{\Delta^{p*}_{75}\Delta^{p}_{57}}(p_0)&
\overline{\Gamma}_{\Delta^{p*}_{75}\Delta^{p*}_{57}}(p_0)&
\overline{\Gamma}_{\Delta^{p*}_{75}\Delta^{p}_{75}}(p_0)&
\overline{\Gamma}_{\Delta^{p*}_{75}\Delta^{p*}_{75}}(p_0)\emat.
\label{19}
\end{equation}
After tedious but straight-forward calculations, based on the
technique elaborated in our previous papers \cite{bekvy,eky1,eky2}
(see also \cite{he,hashimoto}) and used in the consideration of
mesons and diquarks in the color superconducting phase of the two
flavor NJL model, the matrix (\ref{19}) takes the form:
\begin{equation}
\overline{\Gamma}_{57,75}(p_0)=\mat {cccc}
0&A&C&0\\
B&0&0&C\\
C&0&0&A\\
0&C&B&0\emat,
\label{20}
\end{equation}
where $A\equiv\alpha+p_0\beta$, $B\equiv\alpha-p_0\beta$ and 
\begin{eqnarray}
&&\alpha=\int\frac{d^3q}{(2\pi)^3}
\left\{\frac{6E_{\Delta}^+p_0^2-(E_{\Delta}^++E_{2\Delta}^+)^2(
2E_{\Delta}^++E_{2\Delta}^+)}{9E_{\Delta}^+E_{2\Delta}^+[p_0^2-(
E_{\Delta}^++E_{2\Delta}^+)^2]}
+\frac{4p_0^2+4(E_{\Delta}^+)^2-10\Delta^2}
{3E_{\Delta}^+[p_0^2-4(E_{\Delta}^+)^2]}\right\}+\nonumber\\
&&~~~~~~~~~~~~~~~~~~~~~~~+\int\frac{d^3q}{(2\pi)^3}\bigg
\{E_{\Delta}^+\to
E_{\Delta}^-,~~E_{2\Delta}^+\to E_{2\Delta}^-\bigg\},  
\label{21}\\
&&\beta=\int\frac{d^3q}{(2\pi)^3}
\left\{\frac{E^+(E_{\Delta}^++E_{2\Delta}^+)}{3E_{\Delta}^+E_{2\Delta
}^+[p_0^2-(E_{\Delta}^++E_{2\Delta}^+)^2]}+\frac{10E^+}{3E_{\Delta}^+
[p_0^2-4(E_{\Delta}^+)^2]}\right\}-\nonumber\\
&&~~~~~~~~~~~~~~~~~~~~~~-\int\frac{d^3q}{(2\pi)^3}\bigg
\{E^+\to E^-,~~E_{\Delta}^+\to
E_{\Delta}^-,~~E_{2\Delta}^+\to E_{2\Delta}^-\bigg\}, 
\label{22}\\
&&C=\int\frac{d^3q}{(2\pi)^3}
\left\{\frac{2\Delta^2(E_{\Delta}^++E_{2\Delta}^+)}{3E_{\Delta}^+
E_{2\Delta}^+[p_0^2-(E_{\Delta}^++E_{2\Delta}^+)^2]}-
\frac{10\Delta^2}{3E_{\Delta}^+
[p_0^2-4(E_{\Delta}^+)^2]}\right\}+\nonumber\\
&&~~~~~~~~~~~~~~~~~~~~~~+\int\frac{d^3q}{(2\pi)^3}\bigg\{E_{\Delta}^+
\to E_{\Delta}^-,~~E_{2\Delta}^+\to E_{2\Delta}^-\bigg\}.
\label{23}
\end{eqnarray}
(To obtain the above expressions for $\alpha$ and $\beta$, we have
used the gap equation (\ref{15}) in order to eliminate the coupling
constant $G_2$ from corresponding 1PI Green functions.) Evidently,
in the case of mixing between particles the information about their
masses should be extracted from zeros of a matrix determinant,
composed from corresponding 1PI Green functions. So, in our case it
is necessary to study the equation ${\rm det}\overline{\Gamma}
_{57,75}(p_0)=0$, which takes the following form
\begin{equation}
{\rm det}\overline{\Gamma}_{57,75}(p_0)\equiv
(AB-C^2)^2=[(\alpha-C)(\alpha+C)-p_0^2\beta^2]^2=0.
\label{24}
\end{equation}
In the $p_0^2$-plane, each zero of this equation defines a mass
squared of a bosonic excitation of the CFL phase ground state in the
$p(57,75)$ sector. Since this sector contains four pseudoscalar
diquarks, one should search for four solutions of the equation
(\ref{24}) in the $p_0^2$-plane. However, due to the 
structure of (\ref{24}), it is clear that this equation admits at
least two different solutions (each being two-fold degenerate),
which are given by the zeros of the expression in the square
bracket. Since 
\begin{eqnarray}
&&\alpha-C=p_0^2\int\frac{d^3q}{(2\pi)^3}
\left\{\frac{p_0^2[2E_{\Delta}^++4E_{2\Delta}^+]-8(E_{\Delta}^+)^3-
(E_{\Delta}^+)^2E_{2\Delta}^+-10E_{\Delta}^+(E_{2\Delta}^+)^2-
5(E_{2\Delta}^+)^3}{3E_{\Delta}^+E_{2\Delta}^+[p_0^2-(
E_{\Delta}^++E_{2\Delta}^+)^2][p_0^2-4(E_{\Delta}^+)^2]}\right\}+
\nonumber\\&&~~~~~~~~~~~~~~~~~~~~~~~~+p_0^2\int\frac{d^3q}{(2\pi)^3}
\bigg\{E_{\Delta}^+\to E_{\Delta}^-,~~E_{2\Delta}^+\to
E_{2\Delta}^-\bigg\}\equiv p_0^2F(p_0^2),  
\label{25}
\end{eqnarray}
the square bracket in (\ref{24}) becomes zero at the point
$p_0^2=0$. So, in the $p(57,75)$ sector there are two massless
pseudoscalar excitations, i.e. NG bosons. Two other excitations have
the same nontrivial mass squared which is the solution of the
equation 
\begin{equation}
(\alpha+C)F(p_0^2)-\beta^2=0.
\label{26}
\end{equation}
Its investigation is outside the scope of the present paper.
A similar situation occurs in the other four-component sectors
$p(25,52)$ and $p(27,72)$. Namely, for both sectors the 1PI Green
function matrix has the form (\ref{20}). Hence, in each of these
sectors there are two NG bosons as well as two massive excitations
with the same mass squared. Its value is given by the solution of
the equation (\ref{26}).

To summarize, in the pseudoscalar diquark sectors $p(57,75)$,
$p(25,52)$, and $p(27,72)$ we have found six massless diquark
excitations, which are NG bosons, as well as six pseudoscalar
diquark excitations with a nonzero mass which is a solution of the
equation (\ref{26}). In total, these six massive real pseudoscalar
diquarks form a complex (charged) triplet of the SU(3)$_{L+R+c}$
group.

\subsection{The case of the $p(257)$ sector}

All the two-point 1PI Green functions of the pseudoscalar diquarks,
entering the $p(257)$ sector, are defined by the relation
(\ref{17}). In momentum space representation and in the rest frame,
i.e. at $p=(p_0,0,0,0)$, they form the following 6$\times$6 matrix:
\begin{equation}
\overline{\Gamma}_{257}(p_0)=\mat {cccccc}
\overline{\Gamma}_{\Delta^p_{22}\Delta^{p}_{22}}(p_0)&
\overline{\Gamma}_{\Delta^p_{22}\Delta^{p*}_{22}}(p_0)&
\overline{\Gamma}_{\Delta^p_{22}\Delta^{p}_{55}}(p_0)&
\overline{\Gamma}_{\Delta^p_{22}\Delta^{p*}_{55}}(p_0)&
\overline{\Gamma}_{\Delta^p_{22}\Delta^{p}_{77}}(p_0)&
\overline{\Gamma}_{\Delta^p_{22}\Delta^{p*}_{77}}(p_0)\\
\overline{\Gamma}_{\Delta^{p*}_{22}\Delta^{p}_{22}}(p_0)&
\overline{\Gamma}_{\Delta^{p*}_{22}\Delta^{p*}_{22}}(p_0)&
\overline{\Gamma}_{\Delta^{p*}_{22}\Delta^{p}_{55}}(p_0)&
\overline{\Gamma}_{\Delta^{p*}_{22}\Delta^{p*}_{55}}(p_0)&
\overline{\Gamma}_{\Delta^{p*}_{22}\Delta^{p}_{77}}(p_0)&
\overline{\Gamma}_{\Delta^{p*}_{22}\Delta^{p*}_{77}}(p_0)\\
\overline{\Gamma}_{\Delta^p_{55}\Delta^{p}_{22}}(p_0)&
\overline{\Gamma}_{\Delta^p_{55}\Delta^{p*}_{22}}(p_0)&
\overline{\Gamma}_{\Delta^p_{55}\Delta^{p}_{55}}(p_0)&
\overline{\Gamma}_{\Delta^p_{55}\Delta^{p*}_{55}}(p_0)&
\overline{\Gamma}_{\Delta^p_{55}\Delta^{p}_{77}}(p_0)&
\overline{\Gamma}_{\Delta^p_{55}\Delta^{p*}_{77}}(p_0)\\
\overline{\Gamma}_{\Delta^{p*}_{55}\Delta^{p}_{22}}(p_0)&
\overline{\Gamma}_{\Delta^{p*}_{55}\Delta^{p*}_{22}}(p_0)&
\overline{\Gamma}_{\Delta^{p*}_{55}\Delta^{p}_{55}}(p_0)&
\overline{\Gamma}_{\Delta^{p*}_{55}\Delta^{p*}_{55}}(p_0)&
\overline{\Gamma}_{\Delta^{p*}_{55}\Delta^{p}_{77}}(p_0)&
\overline{\Gamma}_{\Delta^{p*}_{55}\Delta^{p*}_{77}}(p_0)\\
\overline{\Gamma}_{\Delta^p_{77}\Delta^{p}_{22}}(p_0)&
\overline{\Gamma}_{\Delta^p_{77}\Delta^{p*}_{22}}(p_0)&
\overline{\Gamma}_{\Delta^p_{77}\Delta^{p}_{55}}(p_0)&
\overline{\Gamma}_{\Delta^p_{77}\Delta^{p*}_{55}}(p_0)&
\overline{\Gamma}_{\Delta^p_{77}\Delta^{p}_{77}}(p_0)&
\overline{\Gamma}_{\Delta^p_{77}\Delta^{p*}_{77}}(p_0)\\
\overline{\Gamma}_{\Delta^{p*}_{77}\Delta^{p}_{22}}(p_0)&
\overline{\Gamma}_{\Delta^{p*}_{77}\Delta^{p*}_{22}}(p_0)&
\overline{\Gamma}_{\Delta^{p*}_{77}\Delta^{p}_{55}}(p_0)&
\overline{\Gamma}_{\Delta^{p*}_{77}\Delta^{p*}_{55}}(p_0)&
\overline{\Gamma}_{\Delta^{p*}_{77}\Delta^{p}_{77}}(p_0)&
\overline{\Gamma}_{\Delta^{p*}_{77}\Delta^{p*}_{77}}(p_0)
\emat.
\label{27}
\end{equation}
A tedious but straight-forward calculation yields:
\begin{equation}
\overline{\Gamma}_{257}(p_0)=\mat {cccccc}
R&P&T&W&T&W\\Q&R&Z&T&Z&T\\T&W&R&P&T&W\\Z&T&Q&R&Z&T\\
T&W&T&W&R&P\\Z&T&Z&T&Q&R\emat.
\label{28}
\end{equation}
The determinant of this matrix looks like
\begin{equation}
{\rm det}\overline{\Gamma}_{257}(p_0)=\big
[(T-R)^2-(W-P)(Z-Q)\big]^2\big\{(2T+R)^2-(2W+P)(2Z+Q)\big\},
\label{29}
\end{equation}
where $P=I_0+p_0I_1$, $Q=I_0-p_0I_1$, $W=J_0+p_0J_1$,
$Z=J_0-p_0J_1$,
and
\begin{eqnarray}
&&I_0(p_0^2)=\frac{1}{4G_2}+\int\frac{d^3q}{(2\pi)^3}
\left\{\frac{28}{9}\cdot\frac{(E_{\Delta}^+)^2+(E^+)^2}{E_{\Delta}^+
[p_0^2-4(E_{\Delta}^+)^2]}
+\frac{4}{9}\cdot\frac{(E_{2\Delta}^+)^2+(E^+)^2}{E_{2\Delta}^+
[p_0^2-4(E_{2\Delta}^+)^2]}+\right.\nonumber\\
&&\left.~~~~~+\frac{2}{9}\cdot\frac{(E_{\Delta}^++E_{2\Delta}^+)[E_{
\Delta}^+E_{2\Delta}^++(E^+)^2]}{E_{\Delta}^+E_{2\Delta}^+
[p_0^2-(E_{\Delta}^++E_{2\Delta}^+)^2]}\right\}+\int\frac{d^3q}{
(2\pi)^3} \bigg\{E_{\Delta,2\Delta}^+\to
E_{\Delta,2\Delta}^-,E^+\to E^-\bigg\},  
\label{30}\\
&&I_1(p_0^2)=\int\frac{d^3q}{(2\pi)^3}
\left\{\frac{28}{9}\cdot\frac{E^+}{E_{\Delta}^+
[p_0^2-4(E_{\Delta}^+)^2]}
+\frac{4}{9}\cdot\frac{E^+}{E_{2\Delta}^+
[p_0^2-4(E_{2\Delta}^+)^2]}+\right.\nonumber\\
&&\left.~~~~~+\frac{2}{9}\cdot\frac{(E_{\Delta}^++E_{2\Delta}^+)E^+}
{E_{\Delta}^+E_{2\Delta}^+
[p_0^2-(E_{\Delta}^++E_{2\Delta}^+)^2]}\right\}+\int\frac{d^3q}{
(2\pi)^3} \bigg\{E_{\Delta,2\Delta}^+\to
E_{\Delta,2\Delta}^-,E^+\to -E^-\bigg\},
\label{31}\\
&&R(p_0^2)=(-\Delta^2)\int\frac{d^3q}{(2\pi)^3}
\left\{\frac{28}{9}\cdot\frac{1}{E_{\Delta}^+
[p_0^2-4(E_{\Delta}^+)^2]}
+\frac{16}{9}\cdot\frac{1}{E_{2\Delta}^+
[p_0^2-4(E_{2\Delta}^+)^2]}~-\right.\nonumber\\
&&~~~~~\left.-\frac{4}{9}\cdot\frac{E_{\Delta}^++E_{2\Delta
}^+}{E_{\Delta}^+E_{2\Delta}^+
[p_0^2-(E_{\Delta}^++E_{2\Delta}^+)^2]}\right\}+\int\frac{d^3q}{
(2\pi)^3} \bigg\{E_{\Delta,2\Delta}^+\to
E_{\Delta,2\Delta}^-\bigg\},
\label{32}\\
&&J_0(p_0^2)=\int\frac{d^3q}{(2\pi)^3}
\left\{\frac{4}{9}\cdot\frac{(E_{2\Delta}^+)^2+(E^+)^2}{E_{2\Delta}^+
[p_0^2-4(E_{2\Delta}^+)^2]}-\frac{2}{9}\cdot\frac{(E_{\Delta}^+)^2+
(E^+)^2}{E_{\Delta}^+[p_0^2-4(E_{\Delta}^+)^2]}~-\right.\nonumber\\
&&\left.~~~~~-\frac{1}{9}\cdot\frac{(E_{\Delta}^++E_{2\Delta}^+)[E_{
\Delta}^+E_{2\Delta}^++(E^+)^2]}{E_{\Delta}^+E_{2\Delta}^+
[p_0^2-(E_{\Delta}^++E_{2\Delta}^+)^2]}\right\}+\int\frac{d^3q}{
(2\pi)^3} \bigg\{E_{\Delta,2\Delta}^+\to
E_{\Delta,2\Delta}^-,E^+\to E^-\bigg\},  
\label{33}\\
&&J_1(p_0^2)=\int\frac{d^3q}{(2\pi)^3}
\left\{\frac{4}{9}\cdot\frac{E^+}{E_{2\Delta}^+
[p_0^2-4(E_{2\Delta}^+)^2]}-\frac{2}{9}\cdot\frac{E^+}{E_{\Delta}^+
[p_0^2-4(E_{\Delta}^+)^2]}~-\right.\nonumber\\
&&\left.~~~~~-\frac{1}{9}\cdot\frac{(E_{\Delta}^++E_{2\Delta}^+)E^+}
{E_{\Delta}^+E_{2\Delta}^+
[p_0^2-(E_{\Delta}^++E_{2\Delta}^+)^2]}\right\}+\int\frac{d^3q}{
(2\pi)^3} \bigg\{E_{\Delta,2\Delta}^+\to
E_{\Delta,2\Delta}^-,E^+\to -E^-\bigg\},
\end{eqnarray}
\begin{eqnarray}
&&T(p_0^2)=(-\Delta^2)\int\frac{d^3q}{(2\pi)^3}
\left\{\frac{16}{9}\cdot\frac{1}{E_{2\Delta}^+
[p_0^2-4(E_{2\Delta}^+)^2]}-\frac{2}{9}\cdot\frac{1}{E_{\Delta}^+
[p_0^2-4(E_{\Delta}^+)^2]}+\right.\nonumber\\
&&~~~~~\left.+\frac{2}{9}\cdot\frac{E_{\Delta}^++E_{2\Delta
}^+}{E_{\Delta}^+E_{2\Delta}^+
[p_0^2-(E_{\Delta}^++E_{2\Delta}^+)^2]}\right\}+\int\frac{d^3q}{
(2\pi)^3} \bigg\{E_{\Delta,2\Delta}^+\to
E_{\Delta,2\Delta}^-\bigg\}.
\label{35}
\end{eqnarray}
In terms of the $I_k,J_l$-functions, the expression (\ref{29}) 
can be rewritten as
\begin{eqnarray}
&&{\rm det}\overline{\Gamma}_{257}(p_0)=\Big
[(T-R-J_0+I_0)(T-R+J_0-I_0)+p_0^2(J_1-I_1)^2\Big]^2\big\{
(2T+R-\nonumber\\
&&~~~~~~~~~-2J_0-I_0)(2T+R+2J_0+I_0)+p_0^2(2J_1+I_1)^2\big\}.
\label{36}
\end{eqnarray}
As in the previous section, the full spectrum of the CFL phase
excitations in the $p(257)$-sector of the model can be explicitly
computed by evaluating the zeros of the determinant (\ref{29}) (or
(\ref{36}), alternatively). Since the $p(257)$-sector consists of
six diquark degrees of freedom, we expect that ${\rm det}\overline
{\Gamma}_{257}(p_0)$ has at least six zeros in the $p_0^2$-plane.

Eliminating in (\ref{30}) the coupling constant $G_2$ again with the
help of the gap equation (\ref{15}), one can easily show that
$T(0)=J_0(0)$ and $R(0)=I_0(0)$. Evidently, in this case both the
square bracket expression and the brace one from (\ref{36}) turn
into zero at $p_0^2=0$. Hence, in the $p(257)$-sector of the model
there are three pseudoscalar NG boson excitations. Now let us obtain
some information about the remaining three excitations, which are
massive ones. For this purpose, similarly to (\ref{25}), it is
possible to extract in an evident form the factor $p_0^2$ from the
expressions in the curly brackets of (\ref{36}), i. e. 
\begin{eqnarray}
&&T-R-J_0+I_0\equiv p_0^2\phi(p_0^2),~~~~~~2T+R-2J_0-I_0\equiv
p_0^2\varphi(p_0^2).
\label{37}
\end{eqnarray}
Then it is clear from (\ref{36}) that two of these excitations have
an identical mass which is a solution of the equation 
\begin{eqnarray}
&&\phi(p_0^2)(T-R+J_0-I_0)+(J_1-I_1)^2=0.
\label{38}
\end{eqnarray}
In total, they form a complex (charged) singlet of the group
SU(3)$_{R+L+c}$. Finally, there is a further diquark excitation,
whose mass obeys another equation 
\begin{eqnarray}
&&\varphi(p_0^2)(2T+R+2J_0+I_0)+(2J_1+I_1)^2=0.
\label{39}
\end{eqnarray}
Evidently, this diquark is also a SU(3)$_{R+L+c}$-singlet.
As a result, we have proved that there are nine pseudoscalar
NG bosons (diquarks) in the mass spectrum of the initial NJL model.
Moreover, it contains massive pseudoscalar diquarks, composed of a
complex triplet and singlet as well as real singlet of the group 
SU(3)$_{f+c}$. \vspace{0.5cm}
 
Since in the chiral limit, where all quarks are massless, the
two-point 1PI Green functions of scalar diquarks are identical to
the corresponding Green functions of pseudoscalar diquark fields,
one may conclude that the diquark spectrum contains nine scalar
massless excitations. In real QCD, eight of them should supply a mass
to gluons through the Anderson -- Higgs mechanism, but the remaining
massless excitation is the NG boson corresponding to the spontaneous
breaking of the baryon U(1)$_B$ symmetry. In addition, there are a
massive complex triplet and singlet as well as a real singlet of
scalar diquarks in the mass spectrum of the model.

\section{Conclusions}

In the present paper the two-point 1PI Green functions of scalar
and pseudoscalar diquarks are investigated in the framework of a
three flavor NJL model with massless quarks and nonzero chemical
potential $\mu$. The model contains interaction terms both in the
quark-antiquark and quark-quark channels, but the `t Hooft six-quark
term is omitted, for simplicity (see (\ref{1})). In this case, the
initial symmetry group of the model, i. e. SU(3)$_L\times$SU(3)$
_R\times$SU(3)$_c\times$U(1)$_B\times$U(1)$_A$ does contain the 
axial U(1)$_A$ subgroup. As a result, we have shown that at
sufficiently low values of $\mu$, $\mu<\mu_c\approx 330$ MeV, the
normal quark matter phase with SU(3)$_{L+R}\times$SU(3)$_c\times
$U(1)$_B$ is realized and nine massless pseudoscalar mesons (which
are the NG bosons), $\pi^{\pm}$, $\pi^0$, $K^0$, $\bar K^0$,
$K^\pm$, $\eta_8$ and $\eta^{\,\prime}$, appear. (In massless QCD,
where U(1)$_A$ is broken on the quantum level, or in NJL models with
`t Hooft interaction the $\eta^{\,\prime}$-meson is not a NG boson.) 

At $\mu>\mu_c$ the original symmetry of the model is spontaneously
broken down to SU(3)$_{L+R+c}$, and the CFL phase does occur. In
this case, in accordance with the Goldstone theorem, eighteen
NG bosons must appear in the mass spectrum. Considering 1PI Green
functions, we have found nine NG bosons in the sector of scalar
diquark excitations. Eight of them have to be considered as
non-physical, since in real QCD they supply masses to gluons by the
Anderson -- Higgs mechanism. The remaining scalar NG boson
corresponds to a spontaneous breaking of the baryon U(1)$_B$
symmetry. The other nine NG bosons are no more pseudoscalar mesons,
but now the massless excitations in the pseudoscalar diquark sector
of the model. 

Besides NG diquarks, we have proved the existence of massive diquark
excitations in the CFL phase. They form both pseudoscalar and scalar
complex (charged) triplets and singlets as well as a real (neutral)
singlet of the group SU(3)$_{f+c}$. 
There arises the interesting question, whether these diquark masses
are above the threshold of fermion pair excitations so that massive
diquarks might eventually decay into two NG bosons. 
\footnote {Our earlier investigations of the dispersion relations of
diquarks in color superconducting quark matter, composed of $u$ and
$d$ quarks (see the papers \cite{eky1}), indicate that massive
diquarks may occur as resonances which are heavily damped. However,
the situation in the CFL phase might be different and needs a special
consideration.}
The detailed numerical investigation of diquark masses as functions
of the chemical potential is outside the scope of this paper and will
be considered in a future publication. There, we are going to study
also the masses of mesons in the environment of dense 
quark matter in the CFL phase. 

\section*{Acknowledgments}

We thank Dr. M. Buballa for useful comments and discussions.
One of us (K.G.K.) gratefully acknowledges the hospitality of the
colleagues at the particle theory group of the Humboldt University
where part of this work was done. This work has been supported in
part by DFG-project 436 RUS 113/477/0-2 and RFBR grant 05-02-16699.

\appendix*
\section{}

In the Nambu -- Gorkov representation the inverse quark propagator
matrix $S_0^{-1}$ is given in (\ref{9}). Using the techniques,
elaborated in \cite{bekvy,eky1,eky2,he,hashimoto}, it is possible to
obtain the following expressions for the matrix elements of the
quark propagator matrix 
$S_0\equiv\vect S_{11},S_{12}\\S_{21},S_{22}\evect$:
 \begin{eqnarray}
\label{121} &&S_{11}(x,y)=\int\!\frac
{d^4q}{(2\pi)^4}\,e^{-iq(x-y)}\left\{
\frac{q_0-E^+}{q_0^2-(E_{B\Delta}^+)^2}
\gamma^0\bar\Lambda_++
\frac{q_0+E^-}{q_0^2-(E^-_{B\Delta})^2}
\gamma^0\bar\Lambda_-\right\},\\
&&S_{12}(x,y)=-i\Delta B\!\int\!\frac{d^4q}{(2\pi)^4}e^{-iq(x-y)}
\left\{\frac{1}{q_0^2-(E_{B\Delta}^+)^2}\gamma^5\bar\Lambda_-+
\frac{1}{q_0^2-(E^-_{B\Delta})^2}\gamma^5\bar\Lambda_+\right\},
\label{122}\\
&&S_{21}(x,y)=-i\Delta^*
B\!\int\!\frac{d^4q}{(2\pi)^4}e^{-iq(x-y)}\left
\{\frac{1}{q_0^2-(E_{B\Delta}^+)^2}\gamma^5\bar\Lambda_++
\frac{1}{q_0^2-(E^-_{B\Delta})^2}\gamma^5\bar\Lambda_-\right\},
\label{123}\\
&&S_{22}(x,y)=\int\!\frac{d^4q}{(2\pi)^4}e^{-iq(x-y)}\left\{
\frac{q_0+E^+}{q_0^2-(E_{B\Delta}^+)^2}
\gamma^0\bar\Lambda_-+
\frac{q_0-E^-}{q_0^2-(E^-_{B\Delta})^2}
\gamma^0\bar\Lambda_+\right\},\label{124}
\end{eqnarray}
where $M=\sqrt{\frac 23}\sigma_0^o$, $\bar\Lambda_\pm=\frac
12(1\pm\frac{\gamma^0(\vec\gamma\vec q- M)}{E})$. Moreover,
$(E_{B\Delta}^\pm)^2=(E^\pm)^2+|\Delta|^2B^2$,
$E^\pm=E\pm\mu$, $E=\sqrt{\strut\vec q^2+M^2}$ and
$B=\sum_{A=2,5,7} \tau_A\lambda_A$. (In these and other
similar expressions, $q_0$ is a shorthand notation for
$q_0+i\varepsilon\cdot {\rm sgn}(q_0)$, where the limit
$\varepsilon\to 0_+$ must be taken at the end of all
calculations. This prescription correctly implements the role of
$\mu$ as 
the chemical potential and preserves the causality
of the theory.)

It is clear from (\ref{121})-(\ref{124}) that all color- and 
flavor dependences in the matrix elements $S_{11},S_{12},S_{21}$ and
$S_{22}$ arise only due to the matrix $B$. In the 
nine-dimensional
space $c\times f$, which is the direct production of color and
flavor spaces, the two $9\times 9$ matrices $B$ and $B^2$ take the
following forms
\begin{equation}
B=\mat {ccccccccc}
0&0&0&0&-1&0&0&0&-1\\0&0&0&1&0&0&0&0&0\\
0&0&0&0&0&0&1&0&0\\0&1&0&0&0&0&0&0&0\\
-1&0&0&0&0&0&0&0&-1\\0&0&0&0&0&0&0&1&0\\
0&0&1&0&0&0&0&0&0\\0&0&0&0&0&1&0&0&0\\
-1&0&0&0&-1&0&0&0&0\emat,~~
B^2=\vect
2,0,0,0,1,0,0,0,1\\0,1,0,0,0,0,0,0,0\\0,0,1,0,0,0,0,0,0\\0,0,0,1,0,0,
0,0,0\\1,0,0,0,2,0,0,0,1\\0,0,0,0,0,1,0,0,0\\0,0,0,0,0,0,1,0,0\\0,0,0
,0,0,0,0,1,0\\1,0,0,0,1,0,0,0,2\evect.
\label{160}
\end{equation}
Thus, the quark propagator $S_0$ may be thought as the matrix in the
$c\times f$ space as well. Since it is an infinite series in powers
of $B$, the diagonalization of the matrix $B$ means the
diagonalization of the propagator $S_0$ in the $c\times f$ space.
Let us define in the $c\times f$ space the following matrix
(its rows are the ortonormal eigenvectors of the $B$-matrix,
corresponding to the eigenvalues $\rho_1=\cdots
=\rho_5=1,\rho_6=\rho_7 =\rho_8=-1,\rho_9=-2$):
\begin{equation}
O=\mat {ccccccccc}
\frac 1{\sqrt{6}}&0&0&0&-\frac 2{\sqrt{6}}&0&0&0&\frac 1{\sqrt{6}}\\
0&\frac 1{\sqrt{2}}&0&\frac 1{\sqrt{2}}&0&0&0&0&0\\
0&0&\frac 1{\sqrt{2}}&0&0&0&\frac 1{\sqrt{2}}&0&0\\
0&0&0&0&0&\frac 1{\sqrt{2}}&0&\frac 1{\sqrt{2}}&0\\
-\frac 1{\sqrt{2}}&0&0&0&0&0&0&0&\frac 1{\sqrt{2}}\\
0&-\frac 1{\sqrt{2}}&0&\frac 1{\sqrt{2}}&0&0&0&0&0\\
0&0&-\frac 1{\sqrt{2}}&0&0&0&\frac 1{\sqrt{2}}&0&0\\
0&0&0&0&0&-\frac 1{\sqrt{2}}&0&\frac 1{\sqrt{2}}&0\\
\frac 1{\sqrt{3}}&0&0&0&\frac 1{\sqrt{3}}&0&0&0&\frac
1{\sqrt{3}}\emat.~~
\label{180}
\end{equation}
Evidently, we have $OO^t=1$, det$O$=1 and
$OB^2O^t=$diag$(1,1,...,1,4)$
$OBO^t=$diag$(1,1,1,1,1,-1,-1,-1,-2)$.
Moreover, the matrix $O$ diagonalizes the quark propagator
$S_0$ in the direct-product space $c\times f$:
$OS_0O^t=$diag$(S_1,...,S_9)$,
where each of $S_1,...,S_8$ corresponds to a Nambu--Gorkov
representation of a propagator of 
the quasiparticle with gap $|\Delta|$, whereas $S_9$ --- 
to a quasiparticle with gap $2|\Delta|$. Hence, in the CLF phase all
nine quasiparticles form an SU(3)-octet with gap $|\Delta|$ and an 
SU(3)-singlet with gap $2|\Delta|$ (see also
\cite{gusynin,buballa}).

\end{document}